\documentclass[epj]{svjour}
\usepackage{graphics}
\begin{document}
\title{Shape transformation transitions in a model of fixed-connectivity surfaces supported by skeletons}

\author{Hiroshi Koibuchi
}                     
%
%
\institute{Department of Mechanical and Systems Engineering \\
  Ibaraki National College of Technology \\
  Nakane 866, Hitachinaka, Ibaraki 312-8508, Japan }
%
%
\abstract{
A compartmentalized surface model of Nambu and Goto is studied on triangulated spherical surfaces by using the canonical Monte Carlo simulation technique. One-dimensional bending energy is defined on the skeletons and at the junctions, and the mechanical strength of the surface is supplied by the one-dimensional bending energy defined on the skeletons and junctions. The compartment size is characterized by the total number $L^\prime$ of bonds between the two-neighboring junctions and is assumed to have values in the range from $L^\prime\!=\!2$ to $L^\prime\!=\!8$ in the simulations, while that of the previously reported model is characterized by $L^\prime\!=\!1$, where all vertices of the triangulated surface are the junctions. Therefore, the model in this paper is considered to be an extension of the previous model in the sense that the previous model is obtained from the model in this paper in the limit of $L^\prime\!\to\!1$. The model in this paper is identical to the Nambu-Goto surface model without curvature energies in the limit of $L^\prime\to \infty$ and hence is expected to be ill-defined at sufficiently large $L^\prime$. One remarkable result obtained in this paper is that the model has a well-defined smooth phase even at relatively large $L^\prime$  just as the previous model of $L^\prime\!\to\!1$. It is also remarkable that the fluctuations of surface in the smooth phase are crucially dependent on $L^\prime$;  we can see no surface fluctuation when $L^\prime\!\leq\!2$, while relatively large fluctuations are seen when $L^\prime\!\geq\!3$.  
}
\authorrunning {H.Koibuchi}
\titlerunning {Shape transformation transitions in a model of fixed-connectivity surfaces supported by skeletons}
\maketitle
\section{Introduction}\label{intro}
One of the interesting topics in membrane physics is to understand why so many varieties of shapes are observed in biological membranes and in artificial membranes \cite{Hotani,Yoshikawa,SEIFERT-LECTURE2004,DS-EPL1996}. A large number of theoretical or numerical studies have been conducted on this problem \cite{NELSON-SMMS2004,Gompper-Schick-PTC-1994,Bowick-PREP2001,Peliti-Leibler-PRL1985,David-Guitter-EPL1988,PKN-PRL1988,SBR-PRA1991,EVANS-BPJ1974,JSWW-PRW1995}. Statistical mechanical view points have shed light on the shape transformations in membranes, and the phenomena are considered to be understood within the theory of phase transitions \cite{GK-SMMS2004}. 

The well-known conventional surface model is the one of Helfrich and Polyakov and defined on triangulated surfaces \cite{HELFRICH-1973,POLYAKOV-NPB1986,KLEINERT-PLB1986}. The Hamiltonian is given by a linear combination of the Gaussian bond potential and the two-dimensional bending energy. The model is known to undergo a collapsing transition and a transition of surface fluctuations on spherical fixed-connectivity surfaces \cite{KANTOR-NELSON-PRA1987,AMBJORN-NPB1993,KOIB-EPJB-2005,KD-PRE2002,KOIB-PRE-2005,KOIB-NPB-2006}. The smooth spherical phase and the collapsed (or crumpled) phase can be seen in the model, however, the above mentioned variety of phases is not expected in the conventional surface model. 

It was recently reported that a multitude of phases can be seen in several surface models \cite{KOIB-PRE2004,KOIB-EPJB2004,KOIB-JSTP-2007,KOIB-EPJB-2007-1,KOIB-EPJB-2007-2,KOIB-EPJB-2007-3,KOIB-PRE2007-2}, which are slightly different from the conventional model and have some additional lattice structures. The surface models that have a multitude of phases can be classified into two groups: The first is a class of Nambu-Goto surface models, where we call a surface model as Nambu-Goto model if the area energy term is included in the Hamiltonian as the bond potential term. In fact, the Nambu-Goto model with intrinsic curvature energy has a variety of phases not only on fixed-connectivity surfaces but also on fluid surfaces \cite{KOIB-PRE2004,KOIB-EPJB-2007-2}. Moreover, the model with one-dimensional bending energy was shown to posses a rich variety of phases \cite{KOIB-EPJB-2007-3}. The second group of models that have a variety of phases is a class of compartmentalized fluid surfaces \cite{KOIB-PRE2007-2,KOIB-EPJB-2006}, where the surface shapes are maintained by a one-dimensional bending energy defined on the compartments. The free diffusion of vertices due to the dynamical triangulations is confined inside the compartments in those models. 

Therefore, the class of Nambu-Goto surface models is quite interesting to study their phase structure as a model for shape transformation phenomena in membranes. As mentioned above, we reported in \cite{KOIB-EPJB-2007-3} that the Nambu-Goto model with the one-dimensional bending energy has a rich variety of phases on triangulated spherical surfaces. However, the model in \cite{KOIB-EPJB-2007-3} has no compartment, because all of the vertices correspond to the junctions and hence, the total number $L^\prime$ of bonds between the junctions is just $L^\prime\!=\!1$. In a surface model with one-dimensional bending energy, one can always assume the compartmentalized structure. Thus, it is natural to extend the model in \cite{KOIB-EPJB-2007-3} by increasing the number $L^\prime$ from $L^\prime\!=\!1$ to non-unital numbers $L^\prime\!>\!1$ so that the model has the compartmentalized structure. 

The notion of well-definedness for surface models should be noted. A triangulated surface model can be called a well-defined one if any physical quantities are finite. The Nambu-Goto model with the conventional two-dimensional bending energy is well-known as an ill-defined model, because the vertices locate in an anomalously extended region in ${\bf R}^3$ and as a consequence, the mean square size $X^2\!=\!\sum_i(X_i\!-\!\bar X)^2$ becomes numerically infinite, where $\bar X$ is the center of mass of the surface. The surfaces are ill-defined also when the relation such as $S_1\!=\!3/2$ is considerably violated, where $S_1\!=\!3/2$ is expected from the scale invariant property in the partition function of the surface model. This ill-definedness comes from the fact that the area energy term imposes a constraint only on the area of triangles, and hence the surfaces are dominated by infinitely thin and infinitely long triangles \cite{ADF-NPB1985}.

The purpose of this paper is to see whether the well-definedness in the model in \cite{KOIB-EPJB-2007-3} is preserved or not when the compartmentalized structure is introduced. It is also aimed at seeing the phase structure. We concentrate on the region of large bending rigidity, where the surface shape is expected to be smooth spherical if the model is well-defined. 

\section{Model and Monte Carlo Technique}
On a triangulated spherical lattice, a sublattice structure is assumed for a discrete Hamiltonian to define the surface model. The triangulated lattices assumed in this paper are identical to those for a compartmentalized fluid surface model in \cite{KOIB-PRE2007-2}. We briefly figure out how to construct the lattice. We start with the icosahedron, which has $N\!=\!12$ vertices of coordination number $q\!=\!5$. Firstly, by dividing the icosahedron uniformly, we have a triangulated lattice of size $N\!=\!10\ell^2\!+\!2$, where $\ell$ is the number of partitions of an edge of the icosahedron. Secondly, a sublattice of size $N_J\!=\!10m^2\!+\!2$, where $m$ divides $\ell$, is assumed on the lattice. Then, all the edges of the sublattice are divided such that the sublattice has the vertices common to the original lattice, and this makes a compartmentalized structure. The total number of vertices $N_S$ on the compartment is given by $N_S\!=\!30m\ell$, and the total number of junctions $N_J$ on the surface is identical to the size $N_J\!=\!10m^2\!+\!2$ of the sublattice \cite{KOIB-PRE2007-2}. Thus, the lattices are characterized by integers $(\ell,m)$ or equivalently by $(N,N_S,N_J)$ and are characterized also by the total number $L^\prime$ of bonds between the junctions. The number $L^\prime$ is given by $L^\prime\!=\!L\!-\!2$, where $L\!=\!\ell/m\!-\!2$, which is identical to that introduced in \cite{KOIB-PRE2007-2}. 

\begin{figure}[hbt]
\centering
\resizebox{0.49\textwidth}{!}{%
\includegraphics{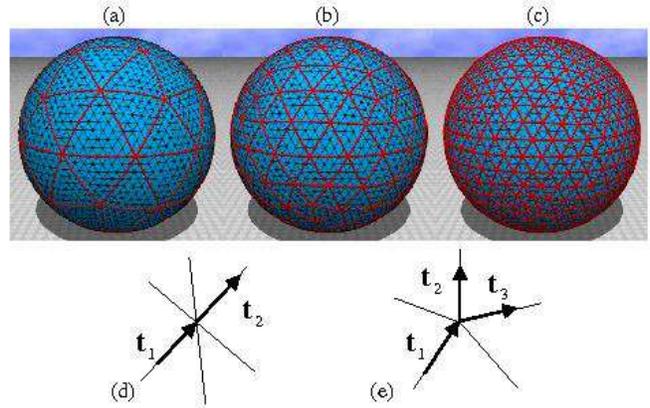}
}
\caption{(Color online) Snapshots of surfaces with skeletons of (a) $L^\prime\!=\!8$, (b) $L^\prime\!=\!4$, and (c) $L^\prime\!=\!2$, whose corresponding integers $(N,N_S,N_J)$ are $(2562, 882, 42)$,  $(2562, 1602, 162)$, and $(2562, 2562, 642)$.  (d) A combination $1\!-\!{\bf t}_1 \cdot {\bf t}_2$ is included in the bending energy with the weight of $1$ at the $q\!=\!6$ junctions,  and (e) a combination $1\!-\!{\bf t}_1 \cdot ({\bf t}_2\!+\!{\bf t}_3)/2$ is included in the bending energy with the weight of $1/2$ at the $q\!=\!5$ junctions. The combination $1\!-\!{\bf t}_1 \cdot {\bf t}_2$ is defined also on the linear chains, which are connected by the junctions such as those shown in (d) and (e). }
\label{fig-1}
\end{figure}
Figures \ref{fig-1}(a)--\ref{fig-1}(c) show the lattices with compartments characterized by the spacing $L^\prime\!=\!8$, $L^\prime\!=\!4$, and $L^\prime\!=\!2$. The corresponding integers $(N,N_S,N_J)$ are given by $(2562, 882, 42)$,  $(2562, 1602, 162)$, and $(2562, 2562, 642)$. Thick lines in the figures are the compartment boundary, which we call {\it skeletons}. The skeletons are composed of the linear chains and the junctions; the linear chains are connected by the junctions. The vertices in the skeletons make the above mentioned sublattice structure.

The Hamiltonian of the model is given by a linear combination of the area energy term $S_1$ and the one-dimensional bending energy $S_2$ such that
\begin{equation}
\label{S1S2}
S=S_1+b S_2, \quad S_1=\sum_{\it \Delta} A_{\it \Delta},\quad S_2=\sum_{ij} (1-{\bf t}_i \cdot {\bf t}_j),
\end{equation}
where $b$ is the bending rigidity. $A_{\it \Delta}$ in $S_1$ is the area of the triangle ${\it \Delta}$, and ${\bf t}_i$ in $S_2$ is a unit tangential vector of the bond $i$. $\sum_{\it \Delta}$ in $S_1$ is the sum over all triangles ${\it \Delta}$, and $\sum_{ij}$ in $S_2$ is the sum over bonds $i$ and $j$ on the skeletons. We should emphasize that $S_2$ is defined only on the skeletons, which are the linear chains and the junctions.

The definition of $S_2$ on the linear chains is straightforward, while it should be remarked at the junctions. The combination $1\!-\!{\bf t}_i \cdot {\bf t}_j$ in $S_2$ at the junctions is defined as follows: Three combinations are defined at the $q\!=\!6$ junctions, and $5/2$ combinations are defined at the $q\!=\!5$ vertices. Figures \ref{fig-1}(d) and \ref{fig-1}(e) show the combinations $1\!-\!{\bf t}_1 \cdot {\bf t}_2$ and $1\!-\!{\bf t}_1 \cdot ({\bf t}_2\!+\!{\bf t}_3)/2$, which are respectively typical of the $q\!=\!6$ vertices and of the $q\!=\!5$ junctions. The combination $1\!-\!{\bf t}_1 \cdot {\bf t}_2$ is included in the summation of $S_2$ with the weight of $1$, while the combination $1\!-\!{\bf t}_1 \cdot ({\bf t}_2\!+\!{\bf t}_3)/2$ is included in $S_2$ with the weight of $1/2$. This is the reason why the $q\!=\!5$ junctions have $5/2$ combinations. 

The statistical mechanical model is defined by the partition function 
\begin{equation} 
\label{Part-Func}
 Z =\int^\prime \prod _{i=1}^{N} d X_i \exp\left[-S(X)\right],
\end{equation} 
where $\int^\prime$ denotes that the multiple integrations should be performed under fixing the center of mass of the surface. The integrations are simulated by the canonical Monte Carlo technique. The vertex position $X_i$ is shifted to a new position $X_i^\prime$ with a three-dimensional random vector $\delta X_i$ such that $X_i^\prime\!=\!X_i\!+\!\delta X_i$. The random vector $\delta X_i$ is chosen in a small sphere, and the radius of the sphere is fixed at the beginning of the simulations so as to have about a $50\%$ acceptance rate; the radius of the sphere depends on $b$ and on whether the surface is in the smooth phase or in the linear phase.

\begin{table}[hbt]
\caption{The surface size assumed in the simulations. Three sizes $(N,N_S,N_J)$ are assumed for each $L^\prime$.}
\label{table-1}
\begin{center}
 \begin{tabular}{cccc}
$L^\prime$  & $(N,N_S,N_J)$ & $(N,N_S,N_J)$  & $(N,N_S,N_J)$  \\
 \hline
  8  & (2562,882,42)   & (5762,1982,92)  & (10242,3522,162)  \\
  6  & (3242,1442,92)  & (5762,2562,162) & (9002,4002,252)   \\
  4  & (2562,1602,162) & (5762,3602,362) & (10242,6402,642)   \\
  3  & (3242,2522,362) & (5762,4482,642) & (9002,7002,1002)   \\
  2  & (2562,2562,642) & (5762,5762,1442) & (10242,10242,2562)   \\
 \hline
 \end{tabular} 
\end{center}
\end{table}
Table \ref{table-1} shows the surface size $(N,N_S,N_J)$ assumed in the simulations in this paper. Three different sizes are assumed for each of the distance $L^\prime$ ranging from $L^\prime\!=\!2$ to $L^\prime\!=\!8$.

The total number of MCS for the thermalization is $2\times10^7\sim 3\times10^7$ in the smooth spherical phase for all surfaces. To the contrary, the speed of convergence is very low in the linear phase especially in the cases $L^\prime\!=\!2$ and $L^\prime\!=\!3$. In fact, the maximum number of the thermalization MCS in the linear phase close to the transition point is about $3\times10^9$ for the surface of $N\!=\!9002$, $L^\prime\!=\!3$, and $2\times10^9$ for the surface of $N\!=\!10242$, $L^\prime\!=\!2$. In the cases of $L^\prime\!=\!4\sim L^\prime\!=\!8$, the thermalization MCS is about $6\times10^8\sim 1\times10^9$ in the linear phase close to the transition point on the largest size surfaces. Relatively smaller number of MCS is done for the thermalization in the smaller sized surfaces in each case of $L^\prime$. The total number of MCS for the production run is $2\times10^8\sim 3\times10^8$ on the smallest surfaces, and $3\times10^8\sim 4\times10^8$ on the middle sized and the largest sized surfaces in each phase and in each $L^\prime$. 

\section{Results}
\begin{figure}[hbt]
\centering
\resizebox{0.46\textwidth}{!}{%
\includegraphics{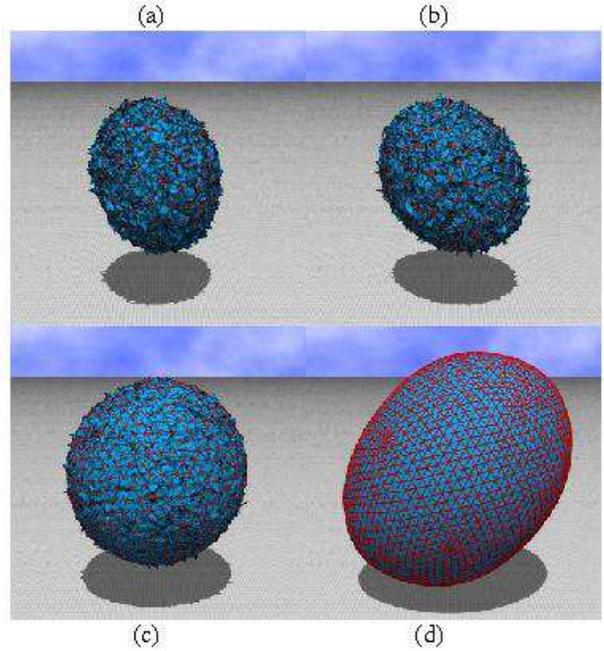}
}
\caption{(Color online) Snapshots of surfaces in the smooth phase obtained at (a) $b\!=\!170$, (b) $b\!=\!220$, (c) $b\!=\!460$, and  (d) $b\!=\!480$, where the surface size are (a) $N\!=\!10242$, $L^\prime\!=\!8$, (b) $N\!=\!9002$, $L^\prime\!=\!6$, (c) $N\!=\!9002$, $L^\prime\!=\!3$, and (d) $N\!=\!10242$, $L^\prime\!=\!2$. Four snapshots were drawn in the same scale.}
\label{fig-2}
\end{figure}
We firstly show in Figs. \ref{fig-2}(a)--\ref{fig-2}(d) snapshots of surfaces in the smooth phase of size (a) $N\!=\!10242$, $L^\prime\!=\!8$, (b) $N\!=\!9002$, $L^\prime\!=\!6$, (c) $N\!=\!9002$, $L^\prime\!=\!3$, and (d) $N\!=\!10242$, $L^\prime\!=\!2$. These snapshots are obtained at (a) $b\!=\!1760$, (b) $b\!=\!220$, (c) $b\!=\!460$, and (d) $b\!=\!480$. These snapshots were drawn in the same scale. From the surface sections, which are not depicted, we see that the surfaces in the smooth phase are empty inside the surface. Thus, we understand that the compartmentalized surface model in this paper has a well-defined smooth phase at sufficiently large $b$ at least when $L^\prime \!\leq \!8$.

\begin{figure}[hbt]
\centering
\resizebox{0.46\textwidth}{!}{%
\includegraphics{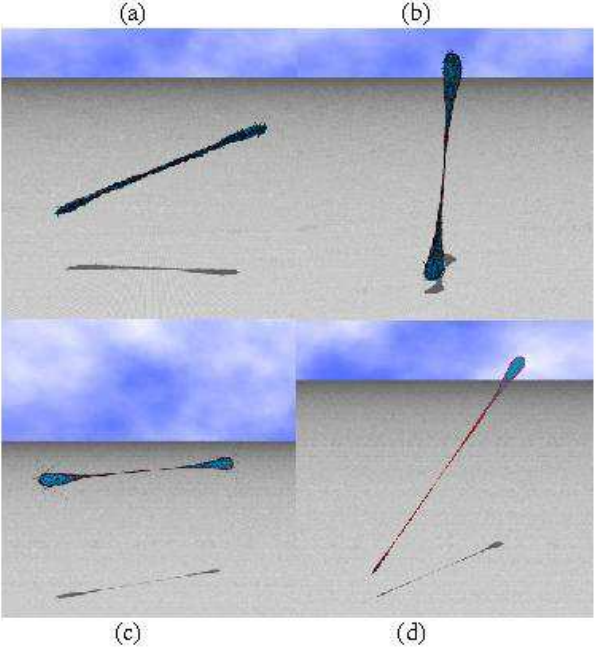}
}
\caption{(Color online) Snapshots of surfaces in the linear phase obtained at (a) $b\!=\!160$, $N\!=\!10242$, $L^\prime\!=\!8$,  (b) $b\!=\!210$, $N\!=\!9002$, $L^\prime\!=\!6$,  (c) $b\!=\!440$,  $N\!=\!9002$, $L^\prime\!=\!3$, and  (d) $b\!=\!460$, $N\!=\!10242$, $L^\prime\!=\!2$, where the surface sizes are identical to those in Figs. \ref{fig-2}(a)--\ref{fig-2}(d), respectively. The scales of the figures are different from each other. }
\label{fig-3}
\end{figure}
Figures \ref{fig-3}(a)--\ref{fig-3}(d) show snapshots in the linear phase at (a) $b\!=\!160$,  (b) $b\!=\!210$, (c) $b\!=\!440$, and  (d) $b\!=\!460$, which are close to the smooth phase in each case. The surface sizes $(N, L)$ in Figs. \ref{fig-3}(a)--\ref{fig-3}(d) are identical with those in  Figs. \ref{fig-2}(a)--\ref{fig-2}(d). The linearly extended surfaces are very different from each other in the length, so the scales of the figures are. We find that the linear phase is also well-defined. The planar phase observed in the model of \cite{KOIB-EPJB-2007-3}, which corresponds to the model in this paper under the condition $L^\prime\!=\!1$, is only seen on the surfaces of $N\!=\!2562$, $L^\prime\!=\!2$ and  $N\!=\!3242$, $L^\prime\!=\!3$. However, the planar phase does not appear on large sized surfaces even when $L^\prime\!=\!2$ or $L^\prime\!=\!3$. We should note that there appear burs or pins on both ends of the surface in Fig. \ref{fig-3}(c) although they are almost invisible. These burs or pins can only be seen on the surfaces of $L^\prime\!=\!3$ in the linear phase close to the transition point, however, the model is considered to be well-defined because of the finiteness of the physical quantities.   

\begin{figure}[hbt]
\centering
\resizebox{0.49\textwidth}{!}{%
\includegraphics{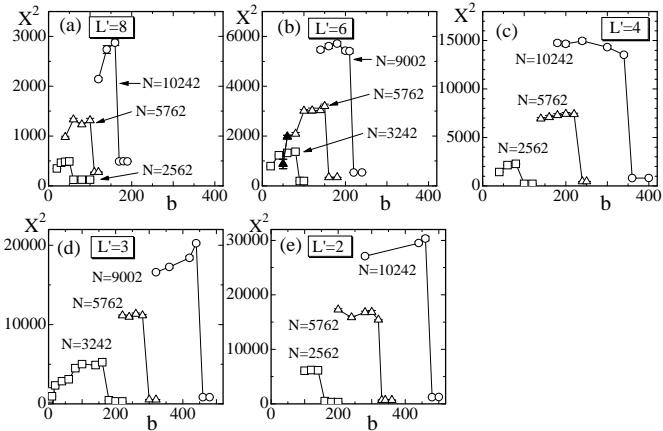}
}
\caption{The mean square size $X^2$ vs. $b$ obtained on the surfaces of size (a) $L^\prime\!=\!8$,  (b) $L^\prime\!=\!6$,  (c) $L^\prime\!=\!4$,  (d) $L^\prime\!=\!3$, and  (e) $L^\prime\!=\!2$. The solid symbols in (b) correspond to ill-defined data, which will be confirmed later in Fig. \ref{fig-8}. }
\label{fig-4}
\end{figure}
The distribution of vertices and hence the surface shape is reflected in the mean square size $X^2$ defined by 
\begin{equation}
\label{X2}
X^2={1\over N} \sum_i \left(X_i-\bar X\right)^2, \quad \bar X={1\over N} \sum_i X_i,
\end{equation}
where ${\bar X}$ is the center of mass of the surface. Figures \ref{fig-4}(a)--\ref{fig-4}(e) show $X^2$ vs. $b$ obtained on the surfaces of $L^\prime\!=\!8$, $L^\prime\!=\!6$, 
 $L^\prime\!=\!4$, $L^\prime\!=\!3$, and $L^\prime\!=\!2$. We should note that the model has no collapsed phase because the model is ill-defined at sufficiently small $b$. The vertices are distributed extremely large volume in ${\bf R}^3$ and behave just like a gas at $b\!\to\! 0$. We know that the model is well-defined in the whole region of $b$ including $b\!\to \!0$ at $L^\prime\!=\!1$, and the well-definedness holds even at  $L^\prime\!\leq\!3$. On the contrary, it is easy to see that the model turns to be ill-defined at $L^\prime\!\geq\!4$ at $b\!\to \!0$. The solid triangular symbols in Fig. \ref{fig-4}(b) correspond to the ill-defined data obtained on the $N\!=\!5762$ surface. This ill-definedness is in fact seen in a large violation of $S_1/N\!=\!3/2$ and is confirmed later. 

The Hausdorff dimension $H$ is defined by $X^2 \sim N^{2/H}$ in the limit of $N\!\to\! \infty$. It is apparently expected that $H_S\!\to\! 2$ in the smooth phase in all cases of $L^\prime$ from $L^\prime\!=\!8$ to $L^\prime\!=\!2$. Then, we obtain $H_L$ in the linear phase by using $X^2$ in that phase at the transition point. Table \ref{table-2} shows $H_S$ in the smooth phase and $H_L$ in the linear phase at the transition point. $H_S$ in Table \ref{table-2} is almost exactly identical to $H_S\!=\! 2$ the expected topological dimension of smooth surfaces, while $H_L$ has values such that $1\!<\! H_L\!<\! 2$. We note that $H_L$ should be $H_L\!=\!1$ if the linear surface has constant density of vertices per unit length. The reason of the deviation of $H_L$ from $H_L\!=\!1$ in Table \ref{table-2} is because the linear surface is not purely one-dimensional.

\begin{table}[hbt]
\caption{The Hausdorff dimensions $H_S$ in the  smooth phase  and  $H_L$ in the linear phase.}
\label{table-2}
\begin{center}
 \begin{tabular}{ccc}
\hline
$L^\prime$ & $H_{\rm S}$    & $H_{\rm L}$   \\
\hline
$8$        & $2.00\pm 0.04$ & $1.57\pm 0.02$  \\
$6$        & $2.00\pm 0.05$ & $1.54\pm 0.11$  \\
$4$        & $2.01\pm 0.04$ & $1.65\pm 0.18$  \\
$3$        & $2.01\pm 0.04$ & $1.50\pm 0.03$  \\
$2$        & $2.00\pm 0.03$ & $1.75\pm 0.09$  \\
 \hline
 \end{tabular} 
\end{center}
\end{table}

\begin{figure}[hbt]
\centering
\resizebox{0.49\textwidth}{!}{%
\includegraphics{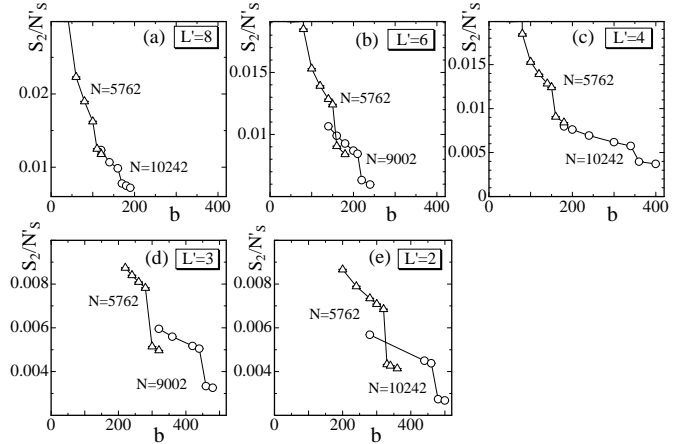}
}
\caption{(a) The one-dimensional bending energy $S_2/N_S^\prime$ vs. $b$  obtained on the surfaces of (a) $L^\prime\!=\!8$,  (b) $L^\prime\!=\!6$,  (c) $L^\prime\!=\!4$,  (d) $L^\prime\!=\!3$, and  (e) $L^\prime\!=\!2$.  }
\label{fig-5}
\end{figure}
Figures \ref{fig-5}(a)--\ref{fig-5}(e) show the bending energy $S_2/N_S^\prime$ vs. $b$ obtained on the surfaces of $L^\prime\!=\!8\sim L^\prime\!=\!2$. The symbol $N_S^\prime$ is given by $N_S^\prime\!=\!N_S\!+\!2N_J\!-\!6$, which can also be written as $N_S^\prime\!=\!10\ell^2\!-\!60m^2\!+\!2$. We see an expected jump in each $S_2/N_S^\prime$ at intermediate $b$, where $X^2$ discontinuously changes.

\begin{figure}[hbt]
\centering
\resizebox{0.49\textwidth}{!}{%
\includegraphics{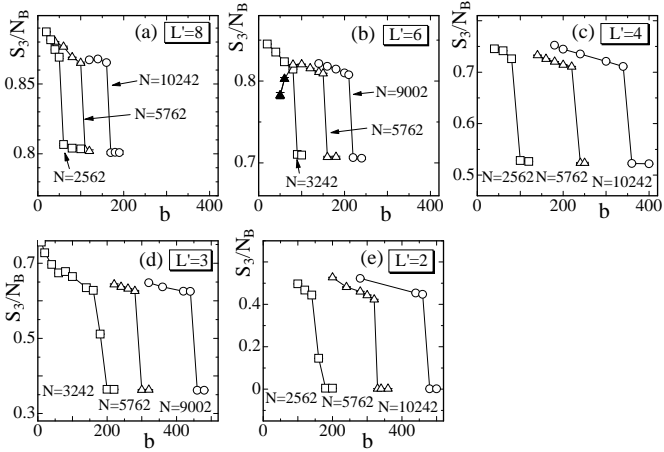}
}
\caption{The two-dimensional bending energy $S_3/N_B$ vs. $b$ obtained on the surfaces of (a) $L^\prime\!=\!8$,  (b) $L^\prime\!=\!6$,  (c) $L^\prime\!=\!4$,  (d) $L^\prime\!=\!3$, and  (e) $L^\prime\!=\!2$. The solid symbols in (b) correspond to ill-defined data, which will be confirmed later in Fig. \ref{fig-8}.}
\label{fig-6}
\end{figure}
The two-dimensional bending energy $S_3$ is defined by
\begin{equation}
\label{S3}
S_3=\sum_{ij} (1-{\bf n}_i \cdot {\bf n}_j),
\end{equation}
where ${\bf n}_i$ is a unit normal vector of the triangle $i$, and $\sum_{ij}$ denotes the summation over all nearest neighbor pairs of the triangles $ij$. $S_3/N_B$ vs. $b$ is plotted in Figs. \ref{fig-6}(a)--\ref{fig-6}(e), where $N_B$ is the total number of bonds including the bonds in the skeletons. Discontinuity in $S_3/N_B$ is very clear just as in the model of \cite{KOIB-EPJB-2007-3}, where $L^\prime\!=\!1$. We see that $S_3/N_B$ is very small in the smooth phase in the case of $L^\prime\!=\!2$. This is because the surfaces are very smooth in that case just like the model in \cite{KOIB-EPJB-2007-3}. Surface fluctuations are almost completely suppressed in the smooth phase when $L^\prime\!=\!2$. This is expected from the compartment structure in the case of $L^\prime\!=\!2$ as shown in the snapshot of Fig. \ref{fig-1}(c). In fact, there is no vertex inside the compartments on the surfaces of $L^\prime\!=\!2$. On the contrary, the total number of vertices in a compartment is greater than one at least in the cases $L^\prime\!\geq\!3$, and no bending energy is assumed on the triangles inside the compartments in those cases. This is the reason why $S_3/N_B$ is relatively large in the smooth phase in the cases $L^\prime\!\geq\!3$. It should also be emphasized that the well-definedness of the surface is remarkable. Only constraint for the vertices and the triangles inside the compartments is the area energy $S_1$ in our model, and no bending energy is assumed on those parts. The solid triangular symbols in Fig. \ref{fig-6}(b) correspond to the ill-defined data obtained on the $N\!=\!5762$ surface.

\begin{figure}[hbt]
\centering
\resizebox{0.42\textwidth}{!}{%
\includegraphics{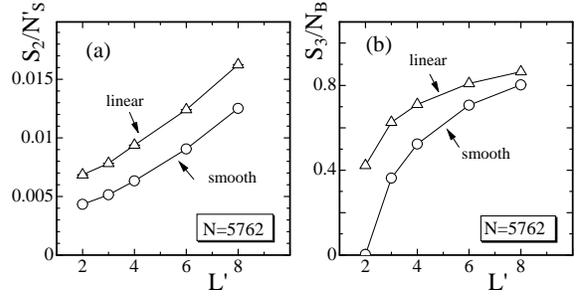}
}
\caption{(a) The bending energy $S_2/N_S^\prime$ vs. $L^\prime$, and (b) the two-dimensional bending energy $S_3/N_B$ vs. $L^\prime$. The surface size is $N\!=\!5762$.}
\label{fig-7}
\end{figure}
We see that the transition point $b_c[kT]$ increases with decreasing $L^\prime$. To see the dependence of physical quantities on $L^\prime$ more clearly, we plot the bending energy $S_2/N_S^\prime$ and the two-dimensional bending energy $S_3/N_B$ against $L^\prime$ in Figs. \ref{fig-7}(a) and \ref{fig-7}(b), where the surface size is assumed as $N\!=\!5762$ for all $L^\prime$. We find from Fig. \ref{fig-7}(a) that $S_2/N_S^\prime$ continuously changes against $L^\prime$. On the contrary, the value of $S_3/N_B$ against $L^\prime$ in the smooth phase appears to be zero (non-zero) when $L^\prime\!\leq\! 2$ ($L^\prime\!>\! 2$). The reason of this behavior in $S_3/N_B$ is because the smoothness of the surface changes almost discontinuously depending on the compartment size $L^\prime$ as mentioned above. 

\begin{figure}[hbt]
\centering
\resizebox{0.49\textwidth}{!}{%
\includegraphics{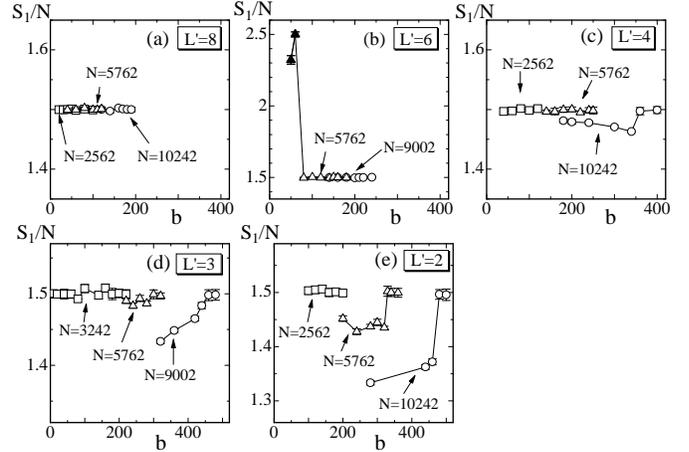}
}
\caption{The Gaussian bond potential $S_1/N$ vs. $b$ obtained on the surfaces of (a) $L^\prime\!=\!8$,  (b) $L^\prime\!=\!6$,  (c) $L^\prime\!=\!4$,  (d) $L^\prime\!=\!3$, and  (e) $L^\prime\!=\!2$. The solid symbols in (b) are obtained at $b\!=\!50$ and $b\!=\!60$ and denote that the model is ill-defined at the rigion of $b \!\leq\! 60$.  }
\label{fig-8}
\end{figure}
Finally, the area energy $S_1/N$ is plotted against $b$ in \ref{fig-8}(a)--\ref{fig-8}(e) in order to see the expected relation $S_1/N\!=\!3/2$ is satisfied. If the surface generated by MC simulations is a well-defined two-dimensional object in ${\bf R}^3$ then $S_1/N\!=\!3/2$ is expected from the scale invariant property of the partition function. We see in the figures that the relation is almost satisfied in the cases $L^\prime\!=\!8$ and  $L^\prime\!=\!6$ except the solid symbols, while the expected relation is slightly violated in the linear phase when $L^\prime\!\leq\!3$. We know the violation of the relation in the model at $L^\prime\!\to\!1$, and therefore the observed violation is consistent to the known result. The reason of a slight violation of the relation is, as described in \cite{KOIB-EPJB-2007-3}, that the vertices move only in a one-dimensional region in ${\bf R}^3$ in the linear phase, and hence the simulation for the three-dimensional integrations is not always well performed on such a linear surface. 

To the contrary, the solid triangular symbols on the $N\!=\!5762$ and $L^\prime\!=\!6$ surface in Fig. \ref{fig-8}(b) clearly represent that the data are ill-defined because $S_1/N$ are far different from $3/2$, where $b\!=\!50$ and $b\!=\!60$. The model is well-defined even at $b\!=\!40$ on the $N\!=\!5762$ and $L^\prime\!=\!8$ surface as we see from Fig. \ref{fig-8}(a), therefore, the lower bound $b$ of well-definedness is not simply dependent only on $L^\prime$. However, the model of $L^\prime \!\geq\! 4$ is ill-defined when $b\!\to\! 0$ in contrast to the cases $L^\prime\! \leq\! 3$, where the model is well-defined even at $b\!\to\! 0$ as mentioned previously in this section.   
 
\section{Summary and Conclusions}
In this paper, we introduced a compartmentalized structure in a Nambu-Goto surface model on triangulated spherical surfaces. The mechanical strength of the surface is given by the one-dimensional bending energy defined only on the compartments with junctions. No bending energy is defined on the triangles inside the compartments, while the area energy term is defined all over the surface. The size of the compartment is characterized by the integer $L^\prime$, which is the total number of bonds between the two neighboring junctions. We previously reported that a large variety of phases can be seen in a Nambu-Goto surface model, which is obtained from the model in this paper in the limit of $L^\prime\!\to\! 1$. In this sense, the model in this paper is an extension of the model in \cite{KOIB-EPJB-2007-3}. Therefore, it was non-trivial whether or not the compartmentalized structure can be introduced into the model in \cite{KOIB-EPJB-2007-3}, because the model might be ill-defined due to the compartments even at sufficiently large bending rigidity $b$. In fact, the model is obviously ill-defined in the limit of $L^\prime\!\to\! \infty$, where neither the compartment nor the bending energy is defined on the surface.

Five different sizes $L^\prime$ of compartments are assumed such that $L^\prime\!=\!8$, $L^\prime\!=\!6$, $L^\prime\!=\!4$, $L^\prime\!=\!3$, and $L^\prime\!=\!2$. For each $L^\prime$, three different sizes $N$ of surfaces are assumed in the range $N\!=\!2562$ $\sim$ $N\!=\!10242$. By using the canonical MC simulation technique we study whether the compartmentalized model is well-defined or not, and moreover we study the phase structure of the model at relatively large $b$ region, where the smooth phase is expected if the model is well-defined. 

Our first observation is that the model is well-defined in the large $b$ region if $L^\prime\!\leq \!8$ at least. In the cases of $L^\prime\!=\!3$ and $L^\prime\!=\!2$, the model is well-defined in the whole region of $b$, while it is not in the small $b$ region when $L^\prime\!\geq \!4$. The second observation is that the model undergoes a first-order shape transformation transition between the smooth spherical phase and the linear phase at those relatively large $b$ region. The planar phase, which is observed between the spherical phase and the linear phase in the model of $L^\prime\!=\!1$ in \cite{KOIB-EPJB-2007-3}, disappears from the compartmentalized model with  $L^\prime\!\geq\!2$. Moreover, no surface fluctuation is observed in the smooth surface in the case of $L^\prime\!=\!2$ just as in the model of \cite{KOIB-EPJB-2007-3}. The surface fluctuations of the smooth spherical surfaces emerge only when $L^\prime\!\geq\!3$ and are completely suppressed when $L^\prime\!\leq\!2$.

\section*{Acknowledgment}
This work is supported in part by a Grant-in-Aid for Scientific Research from Japan Society for the Promotion of Science. 



\end{document}